\documentclass[12pt,a4paper]{article}
\usepackage[dvips]{color,graphicx}
\usepackage{amsfonts}
\usepackage{amssymb}
\usepackage{latexsym}
\usepackage[latin1]{inputenc}

\newcommand{\lw}[1]{\smash{\lower2.ex\hbox{#1}}}

\begin{document}
%
\title{On the meaning of Painlev\'{e}--Gullstrand synchronization}
%
\author{Xavier Ja\'en$^1$\thanks{e-mail address: xavier.jaen@upc.edu} and Alfred Molina$^2$\thanks{e-mail address: alfred.molina@ub.edu}\\
\small $^1$ \parbox[t]{.8\linewidth}{Dept. de F\'isica i Enginyeria Nuclear, Universitat Polit\`ecnica de Catalunya, Spain}\\
\small $^2$ \parbox[t]{.8\linewidth}{Dept. F\'{\i}sica Fonamental, Institut de Ci\`encies del Cosmos, Universitat de Barcelona, Spain}}
\maketitle
%
\begin{abstract}
Following on from two recent papers, here we examine the relationship between Newtonian gravitation and general relativity in more depth. This allows us to define a scalar potential which is just the proper time of the vector potential when the latter is interpreted as the geodesic velocity field. The results are closely related to spacetimes that admit Painlev\'{e}--Gullstrand synchronization.
\end{abstract}
\section{Introduction}
The non-orthogonal coordinate system of Painlev\'e--Gullstrand \cite{Painleve,Morales} is nowadays used to extend the Schwarzschild solution inside its event horizon. The Schwarzschild metric written in this coordinate system is regular across the horizon and is only singular for $r=0$. Another property that is very interesting is that its spatial geometry, the $t=$ constant surfaces, are flat; i.e. what is known as Painlev\'{e}--Gullstrand synchronization.

This type of synchronization is interesting  in the context of gravitational  collapse due to the fact that it allows us to go beyond the Schwarzschild radius. Such synchronization also has  an increasing presence in the literature; for instance, in the so-called analog gravity models \cite{Barcelo} or in relativistic hydrodynamics \cite{Lichne}, where an effective Lorentzian metric is introduced.

Our interest in metrics of this kind stems from the fact that they appear in a natural way when we generalize Newtonian mechanics to general relativity by means of a vector potential that includes gravity and inertial forces on the same footing \cite{Jaen-Molina}. In that earlier work we constructed Newtonian gravitation in which the field is derived from a vector potential  $\vec v_{g}$ that can be interpreted as the velocity field of the trajectory solutions of the equations of motion. Furthermore, the field equations are invariant under the group of rigid motions. This generalization of Newtonian mechanics has a relativistic version, so a significant set of spacetimes can be written in a system of coordinates such that the metric is shape-invariant under the group of rigid motions. Among those spacetimes is the Schwarzschild solution written in Painlev\'{e}--Gullstrand coordinates \cite{Painleve}. Another interesting property of these spacetimes is that their non-relativistic limit can be obtained by making  $ c \rightarrow \infty $  without any consideration regarding weak fields.

In this paper we extend the previous type of metrics introduced in \cite{Jaen-Molina} to a more general one. Besides the vector potential, a scalar potential is introduced that is related to the proper time for the trajectories of free particles. To achieve this generalization, it is very helpful to consider Newtonian gravitation before studying the relativistic case. So first, we analyse Newtonian gravity in relation to the vector potential and the fact that it can be interpreted as a velocity field which is a solution of the equations of motion.

The paper is organized as we explain in what follows. In Sect. \ref{NF}, we study the change pro\-du\-ced in the Newtonian Lagrangian under different choices for the potential velocity field. In Sect. \ref{RF}, we study the behavior of  the relativistic Lagrangian when we change the potential velocity field. In the same section we state the problem and study Minkowski spacetime, introducing vector potentials, in order to express the metric, which are solutions of the equations of motion. As a consequence, we see that we must introduce a new scalar potential. In Sect. \ref{EPGR}, we try to generalize the results of the previous section to a spacetime which depends on four potentials. We see how the properties of these spacetimes are inherited from Newtonian gravitation. In Sect. \ref{PGS}, we see the close relationship between the metrics studied and the family of metrics that support a Painlev\'{e}--Gullstrand synchronization \cite{Bini,Maulik}. In Sect. \ref{RCPN},  we study the form of the metrics under consideration when expressed in adapted coordinates. The results are compared with those obtained by other authors. Finally, in Sect. \ref{FRWL}, we incorporate into this framework the FRWL spacetime  previously considered in \cite{Jaen-Molina-2}.
\section{Newtonian frame\label{NF}}
In \cite{Jaen-Molina} we prove that we can obtain the trajectories of particles in a Newtonian gravity field given by the vector potential $\vec v_g $, from the Lagrangian:
\begin{equation}\label{eq02}
L = \frac{1}{2}m(\dot{\vec x}  - \vec v_g(\vec x,t) )^2 
\end{equation}	
These equations are covariant under rigid motion transformations, i.e. they are the same for inertial and non-inertial frames. If an observer $K$ has a system of orthonormal coordinates $\left\{ {t,\vec x = x^i \vec e_i } \right\}$, and notices a  gravitational field given by the Lagrangian given in (\ref{eq02}),  then observer $K'$ with a system of coordinates that are also orthonormal $\left\{ {t',\vec x' = x'^i \vec e'_i} \right\}$, related to $\left\{ {t,\vec x} \right\}$ through the rigid transformation:
\begin{equation}\label{eq04}
t' = t\;\;;\;\;\vec x = \vec X(t) + {\vec x}{}^\prime = X^i (t)\vec e_i  + x'^i \vec e_i{}'  = X^i (t)\vec e_i  + x'^k R_k^i (t)\vec e_i, 
\end{equation}
where $R_k^i (t)$ is a orthogonal matrix, perceives the same gravitational field (except for inertial forces), i.e. it is produced for the same mass distribution, given by the Lagrangian in (\ref{eq02}), where $\vec x\rightarrow  \vec x{}'$  and $\vec v{}_g\rightarrow  \vec v{}'_g$. The new vector potential   $\vec v{}'_g (\vec x{}',t)$ is related to the old one through:
\begin{equation}\label{eq05}
\vec v_g{}' (\vec x{}',t) = \vec v_g (\vec x,t) - \vec v_0 (\vec x,t)
\end{equation}
where  $\vec v_0 (\vec x,t)$ is the velocity vector field of the trajectories in (\ref{eq04}), i.e.: 
\begin{equation}\label{eq06}
\vec{v_0} (\vec x,t) = \dot{\vec X}(t) + \vec \Omega (t) \times (\vec x - \vec X(t))
\end{equation}
where:	
\begin{equation}\label{eq07}
\vec \Omega (t) \equiv \frac{1}{2}\sum\limits_j {R_j^k (t)} \;\dot R_j^\ell  (t)\;\vec e_k  \times \vec e_\ell  
\end{equation}
It is a trivial and remarkable fact that from the Lagrangian in
(\ref{eq02}), given the vector potential  $\vec v_g (\vec x,t)$, the trajectories that are a solution of $\dot{\vec x}(t) = \vec v_g (\vec x(t),t)$ are a solution  of the equations of motion, i.e. the vector potential  which gives the gravitational field is one of the set of velocity fields which are solutions of the equations of motion of the gravitational field. This result strongly suggests that we can use as a vector potential any other vector potentials $\vec v_g^* (\vec x,t)$ that correspond to another set of trajectories that are also solutions of the equations of motion for the same gravitational field. For instance, for the exterior field of a spherical mass $M$ we have the vector potential:
\begin{equation}\label{eq08}
\vec v_g (\vec x,t) = \sqrt{\frac{{2MG}}
{r}} \;\hat r
\end{equation}
which is the velocity field for the radial trajectories in the gravitational field of the spherical mass $M$ with $v_{\infty}=0$. But what is more interesting is that every velocity field  $\vec v_g^* (\vec x,t)$ corresponding to another set of trajectories that are solutions of the  equations of motion with a potential (\ref{eq08}) is a valid vector potential for the same gravitational field. In the following expression, for each allowed election of the constants $\bar E,\bar \Gamma,\bar \alpha _\phi $, we have a velocity field solution of the equations of motion
\begin{equation}\label{eq09}
\vec v_g^* (\vec x,t) =  \pm \sqrt {\frac{{2MG}}
{r} + 2\bar E - \frac{{\bar \Gamma ^2 }}
{{r^2 }}} \;\hat r\; \pm \frac{1}
{r}\sqrt {\bar \Gamma ^2  - \frac{{\bar \alpha _\phi ^2 }}
{{\sin ^2 \theta }}} \;\hat \theta \; + \frac{1}
{{\;r\sin \theta }}\bar \alpha _\phi  \;\hat \phi \;
\end{equation}
For instance, the vector potential (\ref{eq08}) corresponds to the choice $\bar E=0,\bar \Gamma=0,\bar \alpha _\phi=0 $ in (\ref{eq09}). 
Equation (\ref{eq09}) is a complete family of the velocity fields when the particles move under the potential given by (\ref{eq08}). It is easy to prove that the acceleration field $\vec g$ and the rotation field $\vec \beta$ (see \cite{Jaen-Molina}) are the same when we change the vector potential field $\vec v_g$ (\ref{eq08}) for another $\vec v_g^*$ in the family given by (\ref{eq09}). At the Newtonian level, these are  a kind of gauge transformation.  Let us consider this, from $L$ given in (\ref{eq02}) the momentum is defined as: 
\begin{equation} \label{eq11}
\vec p \equiv \frac{{\partial L}}{{\partial \dot{\vec x}}} =m( \dot{\vec x}  - \vec v_g) \qquad\dot{\vec x} = \vec v_g  + \frac{\vec p}{m}
\end{equation}
and the Hamiltonian is:
\begin{equation} \label{eq12}
\mathcal{H} \equiv \vec p\cdot\dot{\vec x} - \bar L = \frac{{\vec p^2 }}{{2m}} + \vec v_g \cdot\vec p
\end{equation}
The Hamilton-Jacobi equation for the action $S$  is: 
\begin{equation} \label{eq13b} 
\mathcal{H}(\vec x,\vec p = \vec \nabla  S,t) + \frac{\partial S}{\partial t} = 0, 
\end{equation}
or explicitly:
\begin{equation} \label{eq13}
\frac{1}{2m}(\vec \nabla  S)^2  + \vec v_g \cdot\vec \nabla  S + \frac{{\partial  S}}{{\partial t}} = 0
\end{equation}
If we have a solution  $S$ of equation (\ref{eq13}), then  $\vec p = \vec\nabla S $ and from equation (\ref{eq11}) or equivalently from $\dot{\vec{x}}=\frac{\partial\mathcal{H}}{\partial\vec{p}}$, we can construct the trajectory: 
$$\dot{\vec x} = \vec v_g  + \frac{\vec \nabla S}{m}$$
This equation suggests that we can define a new vector potential  (gauge transformation)  $\vec v_g^*  = \vec v_g  + \frac{\vec \nabla S}{m} $ which leads to the same gravitational  and rotational fields, $\vec g $ and $\vec \beta$. The generator of this transformation is the action $S$.

We now show the relation between the Lagrangian $L$ in (\ref{eq02}) and the new one $L^{*}$:
\begin{eqnarray} \label{eq14}
L^{*}  = \frac{m}{2}(\dot {\vec x}  - \vec v_g^* )^2  = \frac{m}{2}\left(\dot {\vec x}  - \left(\vec v_g  + \frac{\vec \nabla S}{m}\right)\right)^2  =  \hfill \nonumber\\ 
  \quad =  L + \frac{1}{2m}(\vec \nabla S)^2  - (\dot {\vec x}  - \vec v_g )\cdot\vec \nabla  S \hfill 
\end{eqnarray} 
The second and third terms on the right-hand side of the last expression, taking into account the  Hamilton-Jacobi equation (\ref{eq13}), can be written as:
\begin{equation} \label{eq15}
\frac{1}{2m}(\vec \nabla  S)^2  - (\dot {\vec x}  - \vec v_g )\cdot\vec \nabla S =  - \frac{{d S}}{{dt}}
\end{equation}
The difference between the two Lagrangians is the total derivative with respect to $t$ of a function, i.e. they are equivalent.
The gauge transformations for the vector potential we have studied produce a change in the non-relativistic Lagrangian which is a total derivative with respect to $t$ of the action; as is well known, this total time derivative does not change the Lagrange equations. This is why in Newtonian mechanics, we have the possibility of writing the field in terms of one vector potential or another; it does not matter. What is interesting is that the set of all possible trajectories of the particles is determined from  knowledge of a single vector potential  which is a particular velocity field of the particle trajectories.
\section{Relativistic frame\label{RF}}
In \cite{Jaen-Molina} we consider the relativistic extension of the Lagrangian in (\ref{eq02}), that is:
\begin{equation}\label{eq34}
L =  - mc^2\sqrt{1-\frac{(\dot{\vec x} - \vec v_g )^2}{c^2}} \,,
\end{equation}

The particle trajectories are the geodesics of the spacetime metric:
\begin{equation}\label{eq35}
ds^2=  -(c^2 -\vec v_g^2 )dt^2+d\vec x^2-2\vec v_g \cdot d\vec x\,  dt
\end{equation}
which has the following properties:
\begin{itemize}
\item it is shape-invariant under the group of rigid motions (\ref{eq04}), provided that the local velocity field transforms as in (\ref{eq05}). (Equation (\ref{eq34}) is a particular case of the so-called {\em Newtonian metrics} \cite{LB}.)
\item in the limit $c\rightarrow\infty$, it gives the Newtonian equations with no need for any weak-field approximation.
\item The integrals of the vector potential $\vec v_g$, i.e. the solutions of $\dot{\vec x}=\vec v_g$, are geodesics.
\end{itemize}
One known example of this type of metric is the Schwarzschild metric written in Painlev\'e--Gullstrand coordinates. But it is clear that to describe all the possibilities  in general relativity (where up to six potentials are needed) this type of metric based on three potentials is not enough. In \cite{Jaen-Molina-2} we add a new potential, $H_g(\vec x,t)$, to describe cosmological metrics such as the FRWL metric. At the end of this paper, in Sect. \ref{FRWL} below, we incorporate FRWL spacetimes. 

The problem with metrics of the form given in equation (\ref{eq35}) is that they are not invariant under the gauge transformations described at the Newtonian level, where the change of $\vec v_g\rightarrow\vec v^*_g$ only changes the Lagrangian by a total derivative of the action with respect to $t$. In the metric given in (\ref{eq35}), when we substitute $\vec v_g$, given in equation (\ref{eq08}) that gives the Schwarzschild metric written in Painlev\'e--Gullstrand coordinates, by $\vec v^*_g$, which should be the relativistic version of equation (\ref{eq09}), the metric is not the Schwarzschild metric anymore. 

Now we will try to identify the problem which appears in the relativistic scheme when we perform the relativistic version of  Newtonian gauge transformations. To this end, we will begin by studying the easiest relativistic spacetime: the Minkowski metric. We write the Minkowski metric for an observer with orthonormal coordinates $\vec x$ and the time is given by watches that are at rest in the $\vec x$ points. 
\begin{equation} \label{eq18}
ds^2  =-c^2 d\lambda ^2  + d\vec x^2 
\end{equation}
$\lambda$ is the proper time for particles at rest. Let us now assume that the observer at the point $\vec x$ does not use the proper time at that point but would like to use the time of  watches  moving following another solution of the equations of motion, for instance a set of watches moving with constant velocity. As we will see, these solutions will give us a velocity field $\vec v^*_g(\vec x,\lambda)$ but also a proper time field   $\tau^*_g(\vec x,\lambda)$.

From the Lagrangian associated with equation (\ref{eq18})\footnote{From now on for simplicity we are going to miss out the $-mc^2$ factor in equation (\ref{eq34})},  $L = \sqrt {1 - \frac{{\dot {\vec x}^2 }}{{c^2 }}}$  where  $\dot {\vec x} = \frac{{d\vec x}}{{d\lambda }}$, we obtain the equation for the action  $S$:
\begin{equation} \label{eq19}
\partial _\lambda  S = \sqrt {1 + c^2 (\vec \nabla S)^2 } 
\end{equation}
Using the expression for the momentum and the inverse relation we have:
\begin{equation}\label{eq20}
\vec p =  - \frac{{\dot {\vec x}}}{{c^2 \sqrt {1 - \frac{{\dot {\vec x}^2 }}{{c^2 }}} }}\, ,\qquad \dot {\vec x}\; = -\frac{{c^2 }}{{\sqrt {1 + c^2 p^2 } }}\vec p
\end{equation}
The velocity equation above suggests that we can construct a vector potential field $\vec v^*_g=\dot{\vec x}$ i.e.:
\begin{equation}\label{eq21}
\vec v_g^{*}  =  - \frac{{c^2 }}{{\sqrt {1 + c^2 (\vec \nabla  S)^2 } }}\vec \nabla  S
\end{equation}
Now we are going to relate the proper time to the action $S$. If $\vec x(\lambda)$ is a solution of the equation $\dot{\vec x}(\lambda)=\vec v^*_g(\vec x(\lambda),\lambda)$, then we are going to see that the proper time  $\tau _g^{*} (\vec x,\lambda )$, is the action $S(\vec x,\lambda)$. Over the trajectories  $\dot{\vec x}(\lambda)=\vec v^*_g(\vec x(\lambda),\lambda)$ we have:
$$d\tau _g^{*}  = \sqrt {1 - \frac{{\vec v_g^{*2}}}{{c^2 }}} \;d\lambda$$
and also:
$$d\tau _g^{*}  = \left( {\partial _\lambda  \tau^*_g  + \vec v_g^{*} \cdot\vec \nabla \tau^*_g } \right)d\lambda$$ 
and the equality on the trajectories of the second member of both equations:
\begin{equation} \label{eq22}
\partial _\lambda  \tau _g^{*}  + \vec v_g^{*} \cdot\vec \nabla \tau _g^{*}  = \sqrt {1 - \frac{{\vec v_g^{*2} }}{{c^2 }}} 
\end{equation}
We are looking for a field $\tau _g^{*}$ that on the trajectories verifies equation (\ref{eq22}). If we have a field that verifies equation (\ref{eq22}) at every point $(\vec x,\lambda)$, then it would also verify equation (\ref{eq22}) on the trajectories. We are going to consider equation (\ref{eq22}) as an equation for $\tau _g^{*}(\vec x,\lambda)$ with $\vec v _g^{*}(\vec x,\lambda) $ known. Then, from equation (\ref{eq21}), we can write equation (\ref{eq22}) as:
\begin{equation} \label{eq23}
\partial _\lambda  \tau _g^{*}  = \frac{{1 + c^2 \vec \nabla  S\cdot\vec \nabla \tau _g^{*} }}{{\sqrt {1 + c^2 (\vec \nabla S)^2 } }}
\end{equation}
and comparing with the Hamilton-Jacobi equation (\ref{eq19}), equation (\ref{eq23}) has the solution for  $\tau^*_g$ :  $\tau _g^{*} (\vec x,\lambda ) =  S(\vec x,\lambda )$ . So the action  $S$ solution of the Hamilton-Jacobi which generates the vector potential  $\vec v_g^*(\vec x,\lambda )$ through equation (\ref{eq21}) can be identified with the proper time of this velocity field. From now on, we identify  $\tau _g^{*}$ and the action  $S$.  This identification has been proposed by other authors, related to a Gaussian coordinate system \cite{Novello}. 

Now it is easy to prove that the metric in (\ref{eq18}) can be written from  $\vec v_g^{*} (\vec x,\lambda )$ and  $\tau _g^{*} (\vec x,\lambda )$ as:
\begin{equation} \label{eq24}
ds^2  =-c^2 d\tau _g^{*2}  + (d\vec x - \vec v_g^{*} d\lambda )^2  + c^2 (\vec \nabla \tau _g^{*} \cdot(d\vec x - \vec v_g^{*} d\lambda ))^2 
\end{equation}
i.e. taking into account equations (\ref{eq19}) and (\ref{eq21}), we can obtain $\partial _\lambda  \tau _g^{*}$ and    $\vec v _g^{*} $: 
\begin{equation} \label{eq25}
\partial _\lambda  \tau _g^{*}  = \sqrt{1+c^2(\vec \nabla \tau _g^{*})^2}\, , \;\;
\vec v_g^{*}  =  - \frac{c^2}{\sqrt{1+c^2(\nabla \tau _g^{*})^2 }} \vec \nabla \tau _g^{*}
\end{equation}
and substituting them in equation (\ref{eq24}), we obtain exactly the same expression (\ref{eq18}).
 
It is important to note that the metric in (\ref{eq24}) may have some ambiguity in the time coordinate used. We can use one of two time coordinates $\tau^*_g$ or $\lambda$. If we do not state otherwise, we will use $\lambda$. This means that $d\tau _g^{*}$ is a short way of writing $d\tau _g^{*}  ={\partial _\lambda  \tau^*_g \;\; d\lambda + d{\vec x} \cdot\vec \nabla \tau^*_g } $ .
It is also important to note that the change of gauge gives exactly the same metric, and therefore the same relativistic Lagrangian. This can be explained because the relativistic origin for the energy is fixed by the rest mass.
\section{Special potentials for general relativity\label{EPGR}}
We are going to construct a spacetime metric invariant for a class of observers in general relativity from a vector potential $\vec v_g(\vec x,\lambda )$ and a proper time $\tau _g(\vec x,\lambda )$. The metric constructed from these fields, following the previous Minkowski example, can be written as:
\begin{equation}\label{metric}
ds^2  = -c^2 d\tau _g^{2}  + (d\vec x - \vec v_g d\lambda )^2 + c^2 (\vec \nabla \tau _g \cdot(d\vec x - \vec v_g d\lambda ))^2 
\end{equation}
Some interesting properties of this kind of metrics are:
\begin{itemize}
\item the metric is shape invariant under transformations where $\lambda$ remains fixed and $\vec x$ transforms as in (\ref{eq04}), i.e. a rigid Newtonian transformation ($\lambda$ represent a kind of absolute Newtonian time) if $\tau_g$ transforms as a scalar function, $\tau '_g (\vec x',\lambda ) = \tau _g (\vec x,\lambda )$, and $\vec v_g (\vec x,\lambda )$  transforms as in (\ref{eq05}) as a Newtonian velocity field. 
\item if we assume that $\tau_g=\lambda+\frac{f(\vec x,\lambda)}{c^2}$, we obtain the Newtonian non-relativistic limit simply by making $c\rightarrow\infty$, with no need for any weak-field approximation.
\item we can define the linear momentum as usual
$$\vec{p}  \equiv {\frac{\partial L}{\partial\dot{\vec x}}} = \frac{1}{L}\left((\partial_\lambda \tau_g+\vec v_{g}\cdot\vec\nabla \tau_g)\vec\nabla \tau_g+\frac{1}{c^2}(\vec v_g-\dot{\vec x})\right)
$$
then $\mathcal{H}\equiv \dot{\vec x}\cdot \vec p-L$
can be written as:
\begin{eqnarray}
\mathcal{H}=&& \vec{v}_g\cdot \vec p+\left(\partial_\lambda \tau_g+\vec v_{g}\cdot\vec\nabla \tau_g\right)\nonumber \\[1ex] &&\left(c^2\vec p\cdot\vec\nabla \tau_g-\sqrt{(1+c^2 \vec{p}^2)(1+c^2 (\vec\nabla \tau_g)^2)}\right)
\end{eqnarray}
and the Hamilton-Jacobi equation is: 
$$
\partial_\lambda S(\vec x,\lambda)+\mathcal{H}(\vec x,\vec p=\vec\nabla S(\vec x,\lambda),\lambda)=0
$$
If we have a particular solution $S$ of this equation, we can obtain the equation of the trajectory: 
\begin{eqnarray}
&&\dot{\vec x}=\left.\frac{\partial \mathcal{H}}{\partial\vec p}\right|_{\vec p=\vec{\nabla} S}= \\[1ex] 
&&\vec{v}_g+c^2(\partial_\lambda \tau_g+\vec{v}_g\cdot\vec{\nabla}\tau_g)\left( \vec{\nabla}\tau_g-\frac{(1+c^2 (\vec{\nabla}\tau_g)^2)\vec{\nabla} S}{\sqrt{(1+c^2 (\vec{\nabla} S)^2)(1+c^2 (\vec{\nabla}\tau_g)^2)}}\right)\nonumber
\end{eqnarray}

It is easy to see that a particular solution of the Hamilton-Jacobi equation is $ S=\tau_g$ and for this solution we have $$\dot{\vec x}=\vec{v}_g$$
Consequently, the trajectory  solutions of $\dot{\vec x}(\lambda)=\vec v_g (\vec x(\lambda),\lambda )$ are solutions of the Lagrange equations i.e. they are geodesics of the metric in (\ref{metric}).

\item It is invariant under the changes between members of the family  of potentials representing different velocity fields and proper times of the  corresponding family of geodesics for the metric in (\ref{metric}). That is, for each particular solution of the Hamilton-Jacobi equation $S$, we can take $\tau^*_g=S$ and ${\vec v_g^{*}}=\left.\frac{\partial \mathcal{H}}{\partial\vec p}\right|_{\vec p=\vec{\nabla} \tau^*_g}$. The metric expressed in these new potentials, $\tau^*_g$ and ${\vec v_g^{*}}$,  is exactly the same as that in (\ref{metric}) and also is written in the same coordinate system.
\end{itemize}
\section{Rigid coordinates and Painlev\'e--Gullstrand synchronization\label{PGS}}
Metric (\ref{metric}) can be expressed in terms of the rigid Newtonian coordinates $\left\{\lambda,\vec x \right\}$ as:
\begin{eqnarray}\label{metricexpl}
 ds^2=&&-c^2 \left(\frac{\partial \tau_g}{\partial\lambda}^2-(\vec\nabla \tau_g \cdot \vec{v}_g)^2-\frac{\vec v_g^2}{c^2}\right)d\lambda^2-\nonumber \\[1ex] && 2 c^2 \left(\frac{\vec v_g}{c^2}+\left(\frac{\partial \tau_g}{\partial\lambda}+\vec v_g\cdot \vec{\nabla} \tau_g\right)\vec\nabla\tau_g\right) d\vec x \; d\lambda + {d\vec x}^2
\end{eqnarray}
This belongs to a family of Newtonian metrics \cite{LB}:
\begin{equation}\label{metricK}
ds^2 = -\Phi(\vec x,\lambda) d\lambda^2+2\vec K(\vec x,\lambda)\cdot d\vec x\,d\lambda+d{\vec x}^2
\end{equation}
the interest in which stems from is based on the existence of the flat slicing $\lambda=\mbox{constant}$ \cite{Morales}; that is, in our rigid Newtonian coordinate system, the spacetime exhibits Painlev\'e--Gullstrand synchronization.

Conversely, given a spacetime metric of the type (\ref{metricK}), we can always construct  the corresponding Hamilton-Jacobi equation: 
\begin{equation}\label{HJK}
\partial_\lambda S= \vec{K}\cdot\vec{\nabla}S+\sqrt{(1+c^2 (\vec{\nabla}S)^2)\left(\left(\frac{K}{c}\right)^2+\frac{\Phi}{c^2} \right)}
\end{equation}
This equation coincides with the scalar equation that we can build considering the equality between the metrics (\ref{metricexpl}) and (\ref{metricK}), by removing $\vec v_g$ and making $\tau_g \rightarrow S$ .

For each particular solution for $S$, from (\ref{HJK}) we obtain a  scalar potential $S_{\mbox{\scriptsize particular}}=\tau_g$. The  vector potential $\vec v_g$ can be obtained from the vector equation linking (\ref{metricexpl}) and (\ref{metricK}):

$$\vec{v}_g=- \vec{K}-c^2 \sqrt{\frac{\left(\frac{K}{c}\right)^2+\frac{\Phi}{c^2}}{1+c^2 (\vec{\nabla} \tau_g)^2} } \;\;\;\;\vec{\nabla} \tau_g$$

Thus, we have demonstrated that given any spacetime that admits a Painlev\'e--Gullstrand synchronization, the Euclidean space coordinates are, in turn, rigid coordinates.

We have also found an interpretation of all Newtonian metrics in terms of a geodesic velocity field and its proper time.
\section{Rigid {\it vs}. Painlev\'e--Novikov\\ adapted to geodesic coordinates\label{RCPN}}
From expression (\ref{metric}), it is clear that the trajectory solutions of 
$$\frac{d\vec x}{d\lambda}=\vec{v}_g(\vec{x}(\lambda),\lambda)$$
are geodesics of proper time $\tau_g$. The general solution of this equation can be expressed as  $\vec{x}=\vec{\varphi}_g(\lambda,\vec y)$, where $\vec y$ is the initial position, i.e. for $\lambda=\lambda_0$ we have $\vec{y}=\vec{\varphi}_g(\lambda_0,\vec y)$. The adapted geodesic coordinates $\left\{\lambda ,\,\vec y\right\}$ can be defined through the relation $\vec{x}=\vec{\varphi}_g(\lambda,\vec y)$. We have:
$$d\vec x=(d\vec y\cdot \vec{\nabla}_y)\vec{\varphi}_g+d\lambda\partial_\lambda\vec{\varphi}_g$$ 
but $\partial_\lambda\vec{\varphi}_g=\vec{v}_g(\vec{\varphi}_g,\lambda)$. Then we have:
$$d\vec x - \vec v_g(\vec{\varphi}_g,\lambda) d\lambda=(d\vec y\cdot \vec{\nabla}_y)\vec{\varphi}_g\equiv \bar d \vec{\varphi}_g $$
where $\bar d f \equiv (d\vec y\cdot \vec{\nabla}_y) f$ is the restriction of the differential form $df$ to each slice, i.e. $d\lambda=0$.

Defining $\tilde \tau _g (\lambda ,\vec y) \equiv \tau _g (\vec x = \vec \varphi _g (\lambda ,\vec y),\lambda )$ we can write: 
$$\vec\nabla_x\tau_g(\vec x,\lambda)\cdot(d\vec x-\vec{v}_g(\vec x,\lambda) d\lambda)|_{\vec x = \vec \varphi _g (\lambda ,\vec y)}=\bar d\tilde\tau_g(\lambda,\vec y)$$
By using the above relations, the metric in (\ref{metric}), in coordinates $\left\{\lambda,\vec{y}\right\}$, becomes: 
\begin{equation}\label{metricadapgeo}
ds^2= -c^2 d{\tilde \tau _g}^2+ c^2 \bar d{\tilde \tau _g}^2 + \bar d \vec \varphi _g^2
\end{equation}
This metric is of the type:
$$ds^2= -c^2 d{\tilde \tau _g}^2+A_{ij}(\vec{y},\lambda) dy^i dy^j$$
which admit as geodesics $\vec y=\mbox{constant}$ with proper time $t=\tilde \tau _g $.

Furthermore, we can use the time coordinate $t$ adapted to the geodesic, i.e. the proper time of the geodesic $\vec y=\mbox{constant}$, through the relation $t=\tilde \tau _g (\lambda ,\vec y)$

Given the potentials, $\tilde \tau_g$ and $ \vec \varphi_g$, in terms of $(\vec y,\lambda )$, we can make the change of time $t = \tilde \tau _g (\vec y,\lambda ) \leftrightarrow \lambda  = \alpha _g (\vec y,t)$ which allows us to write metric (\ref{metricadapgeo}) in the form:
\begin{equation}\label{metricadaptyest}
ds^2  =- c^2 dt^2 + A_{ij} dy^i dy^j 
\end{equation}
meaning that the coordinates  $\left\{ t,\vec y \right\}$ are Gaussian \cite{Novello}. But it is important to note that not all Gaussian coordinates are of the type $\left\{t,\vec y \right\}$, i.e. adapted to a geodesic and rigid. Of course, not all spacetimes will admit a system of coordinates that are adapted to a geodesic and rigid; but all spacetimes admit Gaussian coordinates.

Using the potentials $\alpha_g (\vec y,t)$ and $\vec{\breve{\varphi}}_g(\vec y,t) \equiv \vec \varphi_g (\lambda  = \alpha_g (\vec y,t),\vec y)$ we can write the derivatives $\bar d$  in (\ref{metricadapgeo}), which are restricted to $d\lambda  = 0$ , as: 
$$
\lambda  = \alpha _g (\vec y,t) \Rightarrow d\lambda  = d\vec y\cdot\vec \nabla \alpha _g  + \partial _t \alpha _g dt = 0 \Rightarrow \bar dt =  - \frac{1}{{\partial _t \alpha _g }}d\vec y\cdot\vec \nabla \alpha _g 
$$
then:
$$
\bar d\tau _g  = \bar dt =  - \frac{1}{{\partial _t \alpha _g }}d\vec y\cdot\vec \nabla \alpha_g 
$$
$$
\bar d\vec \varphi_g  = (d\vec y\cdot\vec \nabla )\vec{\breve{\varphi}}_g  + \partial_t \vec{\breve{\varphi}}_g \bar dt = (d\vec y\cdot\vec \nabla )\vec{\breve{\varphi}}_g  - \frac{1}{{\partial_t \alpha_g }}(d\vec y\cdot\vec \nabla \alpha_g )\;\partial_t \vec{\breve{\varphi}}_g 
$$
 The explicit form of (\ref{metricadapgeo}) in adapted coordinates $\left\{ t,\vec y\right\}$ is then: 

\begin{eqnarray}\label{metricadapty}
ds^2  &=& -c^2  dt^2  + c^2 \frac{1}{{(\partial _t \alpha _g )^2 }}(d\vec y\cdot\vec \nabla \alpha _g )^2+\nonumber\\[1ex]
&&\left\{ {(d\vec y\cdot\vec \nabla )\vec{\breve{\varphi}}_g  - \frac{1}{{\partial _t \alpha _g }}(d\vec y\cdot\vec \nabla \alpha _g )\;\partial _t \vec{\breve{\varphi}}_g } \right\}^2 
\end{eqnarray}

Let us now study the special case of spherical symmetry. When we write the metric given in (\ref{metric}) in the rigid coordinate system $(r,\lambda)$, we have:
\begin{eqnarray}
ds^2&=& -c^2d\tau _g(r,\lambda)^{2}+ (dr - v_g(r,\lambda) d\lambda )^2+\nonumber \\[1ex]
&&c^2 (\partial_r\tau_g(r,\lambda)(dr - v_g(r,\lambda) d\lambda ))^2+r^2\, d\Omega^2\nonumber
\end{eqnarray}
The adapted spherical geodesic coordinates $\left\{\lambda, r_0\right\}$ are related to the previous rigid coordinates by:
$$ r=\varphi_g(\lambda,r_0),\quad \mbox{fulfilling}\quad\partial_\lambda\varphi_g=v_g(\varphi_g,\lambda)$$
From (\ref{metricadapgeo}) in the new coordinates $\left\{\lambda, r_0\right\}$, the metric is:
$$ds^2=-c^2 d{\tilde \tau _g}^2+ c^2 \bar d{\tilde \tau _g}^2+\bar d \varphi _g^2+\varphi_g^2\,d\Omega^2$$
or more explicitly: 
\begin{eqnarray}
ds^2&=&-c^2(\partial_{\lambda}\tilde\tau _g\, d\lambda+\partial_{r_0}\tilde\tau _g\, d r_0)^2+\nonumber \\[1ex]
&&\left( ( \partial_{r_0} \varphi_g)^2+c^2 (\partial_{r_0}\tilde\tau _g)^2\right)dr_0^2+\varphi_g^2\,d\Omega^2\label{metricsph}
\end{eqnarray}

Following the same path that led us to (\ref{metricadapty}), we can use the proper time $t$ of the geodesic $r_0=\mbox{constant}$ as the time coordinate, through the relation $t=\tilde \tau _g (\lambda , r=\varphi _g(\lambda,r_0) )\leftrightarrow \lambda  = \alpha_g (r_0 ,t)$, together with $\breve{\varphi}_g (r_0 ,t) = \varphi_g (\lambda  = \alpha_g (r_0 ,t),r_0 )$.  The explicit form of (\ref{metricadapgeo}) in adapted spherical geodesic coordinates $\left\{\lambda, r_0\right\}$ is then: 
\begin{eqnarray}
ds^2&=&-c^2 dt^2  + \left\{c^2 {\left( {\frac{{\partial _{r_0 } \alpha _g }}{{\partial _t \alpha _g }}} \right)^2  + \left( {\frac{{\partial _{r_0 } \alpha _g }}{{\partial _t \alpha _g }}\partial _t \breve{\varphi}_g   - \partial _{r_0 }\breve{\varphi}_g  } \right)^2 } \right\}dr_0 ^2  +\nonumber \\[1ex]
&&\breve{\varphi}_g ^2 \;d\Omega ^2 
\end{eqnarray}

\subsection{Schwarzschild}
If we substitute $\tau_g=\lambda$ and $\vec{v}_g=\sqrt{\frac{2MG}{r}}\hat r$ in (\ref{metric}), we obtain  the Schwarzschild metric written in Painlev\'e--Gullstrand coordinates \cite{Painleve,Jaen-Molina}. We can write this metric in adapted geodesic coordinates. We are going to use the geodesic radial trajectories with radial velocity zero at $r=\infty$ as the new radial coordinate:
\begin{equation}
\tilde\tau_g(r,\lambda)=\lambda,\quad r=\varphi_g(\lambda,r_0)=\left(r_0^{3/2}+\frac32\sqrt{2MG}(\lambda-\lambda_0)\right)^{2/3}\label{solrad}
\end{equation}
Note that in this case, the geodesic proper time $t=\tilde\tau_g$ coincides with $\lambda$. 
The metric written in these coordinates using the expressions (\ref{metricsph}) and (\ref{solrad}) is:
$$ds^2=-c^2d\lambda^2+\frac{r_0}{r} {dr_0}^2+r^2 d\Omega^2$$
where $r$ is given in (\ref{solrad}) as a function of $r_0$ and $\lambda$.  This expression, if we take $\lambda_0=0$ and $r_0=(-\frac32 \sqrt{2MG} \chi)^{2/3}$, coincides with  expression (11.12) of the Schwarzschild exterior solution given by Lema\^{i}tre \cite{Lemetre}, (in the expressions of Lema\^{i}tre $G=1,c=1$). 

The same coordinates are used by Novikov \cite{Novikov}. The relation with our coordinates is $\lambda_0=0$ and $r_0=(\frac32 \xi)^{2/3}$. In expression (3.21),  Novikov used $F=1$. To obtain our expression exactly, he would have needed to take  $F=\sqrt{2GM}$.
\section{Homothetic motions and conformally flat synchronization\label{FRWL}}
Finally, we can incorporate  the FRWL-type  spacetime \cite{Jaen-Molina-2} by considering the metric depending on the  potentials $\vec v_g(\vec x,\lambda)$ , $\tau _g(\vec x,\lambda) $ and $H_g(\vec x,\lambda) ,$:
\begin{equation}\label{metricaFRWL}
ds^2  = -c^2 d\tau _g^2  +c^2 (\vec \nabla \tau _g \cdot(d\vec x - \vec v_g d\lambda ))^2  + \frac{1}{{H_g^2 }}(d\vec x - \vec v_g d\lambda )^2
\end{equation} 

Repeating the same reasoning as in Sect. \ref{EPGR} step by step, where $H_g  = 1$,  we conclude that:

\begin{itemize}
\item  metric (\ref{metricaFRWL}) is shape invariant under the group of  homothetic motions $\left\{{X^i(\lambda),R_j^i(\lambda),H(\lambda)} \right\}$, which generalize the group of rigid motions $\left\{{X^i(\lambda),R_j^i(\lambda)} \right\}$ \cite{Jaen-Molina-2}, if the vector potential  $\vec v_g(\vec x,\lambda)$ and $H_g(\vec x,\lambda)$ transform as $\vec v'_g(\vec x',\lambda)$ and $H'_g(\vec x',\lambda)$:
\begin{equation}
\vec v'_g(\vec x',\lambda) = \vec v_g(\vec x,\lambda)  - \vec v_0(\vec x,\lambda); \;\;\;H'_g(\vec x',\lambda)  = H(\lambda) H_g(\vec x,\lambda) ,
\end{equation}
where $\vec v_0$ is now defined, not as in (\ref{eq06}), but according to:
\begin{equation}
\vec v_0 (\vec x,\lambda) = \vec V(\lambda) + \vec \Omega (\lambda) \times (\vec x - \vec X(\lambda)) + \frac{{\dot H(\lambda)}}
{H(\lambda)}(\vec x - \vec X(\lambda)),
\end{equation}
where $\vec \Omega$ is defined in (\ref{eq07}) and ``$\times$" is the ordinary vector product; and $\tau_g$ transforms as a scalar. 
\item if $\tau_g=\lambda+\frac{f(\vec x,\lambda)}{c^2}$ and $H_g(\vec x,\lambda)=H_g(\lambda)+\frac{F(\vec x,\lambda)}{c^2}$, then we obtain the Newtonian non-relativistic limit simply by making $c\rightarrow\infty$, with no need for any weak-field approximation.
\item we can define the linear momentum as usual:
$$\vec{p}  \equiv {\frac{\partial L}{\partial\dot{\vec x}}} = \frac{1}{L}\left((\partial_\lambda \tau_g+\vec v_{g}\cdot\vec\nabla \tau_g)\vec\nabla \tau_g+\frac{1}{c^2 H_g^2}(\vec v_g-\dot{\vec x})\right)
$$
and then $\mathcal{H}\equiv \dot{\vec x}\cdot \vec p-L$
can be written as:
\begin{eqnarray}
\mathcal{H}&=&\vec{v}_g\cdot \vec p+(\partial_\lambda \tau_g+\vec v_{g}\cdot\vec\nabla \tau_g)\nonumber \\[1ex] &&\left(c^2 H_g^2 \; \vec p\cdot\vec\nabla \tau_g-\sqrt{(1+c^2 H_g^2 \;  \vec{p}^2)(1+c^2 H_g^2 (\vec\nabla \tau_g)^2)}\right)
\end{eqnarray}
and the Hamilton-Jacobi equation is: 
$$
\partial_\lambda S(\vec x,\lambda)+\mathcal{H}(\vec x,\vec p=\vec\nabla S(\vec x,\lambda),\lambda)=0
$$
If we have a particular solution of this equation, we can obtain the trajectory equation: 
\begin{eqnarray}
\dot{\vec x}&=& \left.\frac{\partial \mathcal{H}}{\partial\vec p}\right|_{\vec p=\vec{\nabla} S}  
=\vec{v}_g+c^2 H_g^2 \left(\partial_\lambda \tau_g+\vec{v}_g\cdot\vec{\nabla}\tau_g\right) \nonumber \\[1ex] &&\left( \vec{\nabla}\tau_g-\frac{(1+c^2  H_g^2 (\vec{\nabla}\tau_g)^2)\vec{\nabla} S}{\sqrt{(1+c^2 H_g^2 (\vec{\nabla} S)^2)(1+c^2 H_g^2 (\vec{\nabla}\tau_g)^2)}}\right)
\end{eqnarray}

It is easy to see that a particular solution of the Hamilton-Jacobi equation is $ S=\tau_g$ and for this solution we have: $$\dot{\vec x}=\vec{v}_g$$
Consequently, the trajectory  solutions of $\dot{\vec x}(\lambda)=\vec v_g (\vec x(\lambda),\lambda )$ are solutions of the Lagrange equations i.e. are geodesics of the metric in (\ref{metricaFRWL}).

\item It is invariant under the changes between members of the family  of potentials, $\vec v_g$ and $\tau_g$, which represent different velocity fields and proper times of the  corresponding family of geodesics for (\ref{metricaFRWL}). That is, for each particular solution of the Hamilton-Jacobi equation $S$ we can take $\tau^*_g=S$ and ${\vec v_g^{*}}=\left.\frac{\partial \mathcal{H}}{\partial\vec p}\right|_{\vec p=\vec{\nabla} S}$. The metric related to these new potentials, $\tau^*_g$ and ${\vec v_g^{*}}$, now with $H_g$ unchanged, is exactly the same as in equation (\ref{metricaFRWL}).

\item The results of the slicing $\lambda=$constant are now conformally flat $ds^2  =  \frac{1}{{H_g^2 }}d\vec x^2 $. Metric (\ref{metricaFRWL}) belongs to a family of metrics that admit a conformally flat synchronization 

\begin{equation}\label{metricaFRWLK}
ds^2= -\Phi(\vec x,\lambda) d\lambda^2+2\vec K(\vec x,\lambda)\cdot d\vec x\,d\lambda+\frac{1}{{H_g^2 }}d{\vec x}^2
\end{equation}

Conversely, given a spacetime metric of type (\ref{metricaFRWLK}), we can always construct the corresponding Hamilton-Jacobi equation:

\begin{equation}\label{HJRWLK}
\partial_\lambda S= H_g^2 \vec{K}\cdot\vec{\nabla}S+\sqrt{(1+c^2 H_g^2 (\vec{\nabla}S)^2)\left(H_g^2 \left(\frac{K}{c}\right)^2+\frac{\Phi}{c^2} \right)} 
\end{equation}

This equation coincides with the scalar equation that we can construct considering the equality between the metrics (\ref{metricaFRWL}) and (\ref{metricaFRWLK}), by removing $\vec v_g$ and making $\tau_g \rightarrow S$ .

For each particular solution $S$ of equation (\ref{HJRWLK}), we obtain a  scalar potential  $\tau_g = S$. The  vector potential $\vec v_g$ can be obtained from the vector equation linking (\ref{metricaFRWL}) and (\ref{metricaFRWLK}) or directly from the Hamiltonian.

$$\vec{v}_g=-H_g^2\left(\vec{K}+ c^2\sqrt{\frac{H_g^2\left(\frac{K}{c}\right)^2+\frac{\Phi}{c^2}}{1+c^2 H_g^2(\vec{\nabla} \tau_g)^2} }\vec{\nabla} \tau_g\right)$$ 

Thus, we have demonstrated that given any spacetime that supports a conformally flat synchronization, its conformal Euclidean coordinates are, in turn, homothetic coordinates.
\end{itemize}
\section{Conclusions}
This work is a continuation of two previous papers. In that previous work, we analysed some almost unknown properties of Newtonian gravity. We then tried to translated those properties to relativistic gravity. This approach allowed us, in the first paper, to provide a meaning for a family of metrics, which are nothing more than an extension of the Painlev\'e--Gullstrand type of metrics. In the second paper we were able to incorporate cosmological spacetimes, of the FRWL type, into a framework that can be considered a cosmological extension of the Painlev\'e--Gullstrand type of metrics.

In this way, we were able to find up to four potentials, which have their own meaning, from which to express a significant set of spacetime metrics. 

It is clear that the set of spacetimes covered by the metrics we had found, with only four potentials, was not enough to include the set covered by standard general relativity. But the fact that the study of Newtonian gravity was so  successful encouraged us to extend our research even further in the same direction.

To this end, in the present  paper, we develop a new unknown property of Newtonian gravity; a property that already appeared as a curiosity in our previous work: the vector potential can be interpreted as a velocity field whose integral trajectories are solutions of the equations of motion.

This fact strongly suggests that the vector potential can be substituted by any other vector potential that is also a solution of the equations of motion. And that is indeed the case at the non-relativistic level. The field equations are invariant under this kind of gauge transformation.

The main contribution of this paper is to extend this property to the maximum number of spacetimes within standard general relativity. We clearly laid out the problem and we saw that we needed a new scalar potential whose meaning is the proper time of the vector potential. Studying the Minkowski metric allowed us to construct a sufficiently general family of metrics which inherit the properties of Newtonian gravity above mentioned.

The family of metrics found, expressed in a rigid coordinate system, is the same as the fa\-mi\-ly of metrics with flat synchronization, i.e. those that exhibit Painlev\'e--Gullstrand synchronization.

Finally, we incorporated the cosmological spacetime into the general fra\-me\-work, without losing any of the properties inherited from Newtonian gravity. Now the family of metrics, expressed in homothetic coordinates, is the same as the family of metrics with conformally flat  synchronization.

As a result, we obtain a family of spacetimes which support a conformally flat synchronization; the metric can be written via five potentials:
\begin{itemize}
\item Three components of the vector potential ${\vec v}_g$, plus one of the scalar potential ${\tau}_g$, from which we can form a geodesic four-vector of the metric which in turn it represents.
\item A scalar potential $H_g$ which has the meaning of a local homothety.
\end{itemize}
All these potentials have some physical meaning, and even in Newtonian theory they leave a trace.

As is well known, the general theory of relativity is described by six potentials. In our approach, in which we have been guided by Newtonian theory, we have identified five of these six potentials. An interesting question is: is it possible to introduce the sixth potential without sacrificing the properties studied in this work? Then, if this is not possible: what are the properties we should dispense with?

The spacetimes covered in this work are identified with those that admit a flat or conformally flat synchronization. These spacetimes do not support gravitational waves.  It seems that the introduction of the sixth potential will allow gravitational waves. Following these comments, it seems clear that the introduction of the sixth potential cannot arise from any previous study of Newtonian gravitation or of special relativity. The sixth potential may be most genuinely linked to general relativity, without leaving any trace possible of being included in Newtonian theory by some limiting process.
\section*{Acknowledgments}
We thank Bartolomé Coll for suggestions we received at an early stage of the work; Juan Antonio Morales-Lladosa and Eduardo Ruiz for carefully reading a previous draft of the paper and providing useful criticism that led to improvements; and finally Llu\'{\i}s Bel, without whose inspiration and encouragement, hardly any of this series of papers would have occurred to us.


\begin{thebibliography}{99} 
\bibitem{Painleve} Painlev\'e, P.  {\it  C.R. Acad. Sci. (Paris)}{\bf 173}, 677-680 (1921). Gullstrand, A. {\it Ark. Mat. Astron. Fys.}, {\bf 16} (8), 1-15 (1922).
\bibitem{Morales} Herrero, A. and Morales,J.A. {\it Class. Quantum. Grav.} {\bf 27} 175007, (2010) 
\bibitem{Barcelo} Barcel\'o, C., Liberati, S. and Visser, M. {\it Living Rev. Relativity} {\bf 14} 3, (2011). DOI 10.12942/lrr-2011-3.
\bibitem{Lichne} Lichnerovicz, A. {\it Relativistic Hydrodynamics and Magnetohydrodynamics} Benjamin, (1967).
\bibitem{Jaen-Molina} Ja\'en, X. and Molina, A. {\it Gen. Relativ. Gravit.}{\bf 45}, 1531-1546, (2013) DOI 10.1007$_s$10714-013-1542-9.
\bibitem{Bini} Bini, D. Geralico, A. and Jantzen, R.T. {\it Gen. Relativ. Gravit.}{\bf 44}, 603-621, (2012) DOI 10.1007$_s$10714-011-1295-2.
\bibitem{Maulik} Maulik K. Parikh, {\it Physics Letters B} {\bf 546} 189-195 (2002).
\bibitem{Jaen-Molina-2} Ja\'en, X. and Molina, A. {\it Gen. Relativ. Gravit.}{\bf 46}, 1745, (2014) DOI 10.1007$_s$10714-014-1745-8.
\bibitem{LB} Bel, Ll. {\it Recent developments in gravitation} Ed. Verdaguer, Garriga, Cespedes World Scientific (1990).
\bibitem{Novello} Novello, M. and Bittencourt, E. {\it Gravitation and Cosmology} {\bf 17} (3), 230-241 (2011).
\bibitem{Lemetre} Lema\^{i}tre, G. {\it Annales de la Soci\'{e}t\'{e} Scientifique de Bruxelles} {\bf A53} 51-85 {1993}. Reproduced in
{\it Gen. Relativ. Gravit.}{\bf 29}, 641-680 (1997).
\bibitem{Novikov} Novikov, I. D. {\it Communications of the Shternberg State Astronomical Institute} {\bf 132} 3-42 (1964). Reproduced in {\it Gen. Relativ. Gravit.} {\bf 33} 2259-2295 (2001).
\end{thebibliography}
\end{document}